\documentclass[%
 reprint,
groupedaddress,
 amsmath,amssymb,
 aps,
prb,
]{revtex4-1}

\usepackage{graphicx}
\usepackage{dcolumn}
\usepackage{bm}

\begin{document}

\preprint{APS/123-QED}

\title{Non-Rayleigh distribution of reflected intensity from photonic crystals with disorder}

\author{Oluwafemi S. Ojambati}\email{o.s.ojambati@utwente.nl}
\author{Elahe Yeganegi} 
\altaffiliation{Current address: ASML, Flight Forum 1900, 5657 EZ, Eindhoven, The Netherlands
}
\author{Ad Lagendijk}
\author{Allard P. Mosk} \altaffiliation{Current address: Physics of Light in Complex Systems, Debye Institute for Nanomaterials Science, Utrecht University.
}
\author{Willem L. Vos}
\affiliation {
Complex Photonic Systems (COPS), MESA+ Institute for Nanotechnology, University of Twente, \\
P.O. Box 217, 7500 AE Enschede, The Netherlands}

\date{\today}

\begin{abstract}
Structural disorder results in multiple scattering in real photonic crystals, which have been widely used for applications and studied for fundamental interests.
The interaction of light with such complex photonic media is expected to show interplay  between disorder and order. 
For a completely disordered medium, the intensity statistics is well-known to obey Rayleigh statistics with a negative exponential distribution function, corresponding to absence of correlations.
Intensity statistics is unexplored however for complex media with both order and disorder.
We study experimentally the intensity statistics of light reflected from photonic crystals with various degree of disorder. 
We observe deviations from the Rayleigh distribution and the deviations increase with increasing long-range order. 
\end{abstract}

\pacs{Valid PACS appear here}
\maketitle


\section{Introduction}
{\renewcommand{\thefootnote}{}
\footnotetext{In preparation as O.S. Ojambati, E. Yeganegi, A. Lagendijk, A.P Mosk, and W.L. Vos.}
}

Mesoscopic transport of waves through disordered scattering media has been widely studied for interesting fundamental phenomena~\cite{Anderson1958PRL,Albada1985PRL,Wolf1985PRL,Shapiro1986PRL,Feng1988PRL,Berkovits1994PhysRep,scheffold1998PRL,vanRossum1999RevModPhys} and exciting applications~\cite{leith1966JOSA,Dowling1992JourAcouAm,wiersma1996PRE,Vynck2012NatMat,Redding2013NatPhoton}.
In a disordered medium, multiple scattering of waves scrambles the amplitudes and phases of the waves. 
Random interference occurs between the multiply scattered coherent waves and results in the well-known speckles, which are observed as rapid fluctuations of high and low intensities~\cite{Goodman2007SpeckleBook,Goodman2000StatOptBook}. 


Another interference effect of multiply scattered waves in disordered media is mesoscopic correlation~\cite{Watson1969JMathPhys,Golubentsev1984SovPhysJETP}.
It is well-known in quantum transport theory for electrons -- also applicable to classical waves such as light and sound -- that mesoscopic correlations originate from the crossing of wave trajectories inside the medium~\cite{akkermans2007book}.
To study mesoscopic correlations, one of the main statistical tools that have been widely used is intensity histogram that is compared with Rayleigh distribution: a negative exponential function~\cite{Garcia1989PRL,Shnerb1991PRB,deBoer1994PRL,Chabanov2000Nature,Hu2008NatPhys,Strudley2013NatPhoton,Strudley2014OptLett,Bromberg2014PRL,Zhou2015OptExp}.
The Rayleigh distribution is expected for a field that is a sum of waves with statistically independent and uniformly distributed amplitudes and phases~\cite{Goodman2007SpeckleBook,Goodman2000StatOptBook}.
The intensity transmitted through a large ensemble of disordered scatterers is known to follow the Rayleigh distribution to good approximation~\cite{Garcia1989PRL}.
An interesting scenario is when the intensity distribution deviates from the Rayleigh distribution.
Such a deviation in a completely disordered system can be an indicator of mesoscopic correlations, while in a partially ordered system it is a sign of the interplay between order and disorder~\cite{Zhang2002PRB}.

To quantify deviations from the Rayleigh distribution, speckle contrast (SC) is a useful parameter. It is defined as $\rm{SC} \equiv \sigma/\langle I \rangle $, where $\sigma$ and $\langle I \rangle $ are the standard deviation and mean, respectively, of the intensity.
When the intensity histogram follows the Rayleigh distribution, we find SC = 1 ~\cite{Goodman2000StatOptBook}. 
Two scenarios are known to lead to $\rm{SC} > 1$: (i) when the phases and amplitudes are not uniformly distributed, for example, in the case of partially developed speckle due to weak scattering with significant contribution of ballistic light~\cite{GMartin_PRL_2000}, and (ii) when the field is a random phasor sum of waves with crossing trajectories, which leads to mesoscopic correlation~\cite{Watson1969JMathPhys,Golubentsev1984SovPhysJETP}.
These two scenarios of non-Rayleigh statistics could be juxtaposed if the scattering properties of the sample is known \textit{a priori}.

Non-Rayleigh distributions have been observed experimentally for several systems: Collections of a small number of scatterers~\cite{Schaefer1972PRL}, a rough surface~\cite{Jakeman1973JPhysA,Ohtsubo1978OptComm}, the Anderson localization regime~\cite{Chabanov2000Nature}, anisotropic scattering in disordered mats of semiconductor nanowires~\cite{Strudley2013NatPhoton}, isotropic scattering in a large disordered ensemble of nanoparticles~\cite{Strudley2014OptLett}, and tailored incident waves~\cite{Bromberg2014PRL}.
In all these observations, disordered scattering media were studied.
An interesting but unexplored area is the intensity statistics of samples with both disorder and order such as a real photonic crystal. 
In a perfect crystal, all intensities due to Bloch waves are completely correlated as they are given by the crystal structure factors~\cite{AshcrofBook,Warren1969Book}.
In addition to Bloch waves, there is a contribution of uncorrelated scattered waves due to disorder in a real photonic crystal. 
A fundamental question addressed here is if the correlation is completely destroyed due to the influence of disorder in photonic crystals.

In this paper, we present the first experimental observation of non-Rayleigh distribution of reflected intensities from photonic crystals with inevitable fabrication-induced disorder.
The samples investigated here are self-assembled artificial 3D opals of silica spheres and 2D Si lithographically-etched photonic crystals. 
We observe significant deviations from the Rayleigh distribution for light reflected from photonic crystals, depending on  the amount of disorder.
This deviation is rather surprising since these photonic crystals are considered to be strongly scattering with a transport mean free path of the order of 10$~\mu$m~\cite{Koenderink2005PRB,Muskens2011PRB}.
We attribute the non-Rayleigh distribution to an interplay of the underlying structural correlation with disorder. 
Our results could also be applied to  media with intentional correlated disorder such as hyper-uniform structures~\cite{Florescu2009PNAS,Muller2014AdvOptMater}.


\section{Experimental setup and samples}

\begin{figure}[h]
\center
\includegraphics[width=0.45\textwidth]{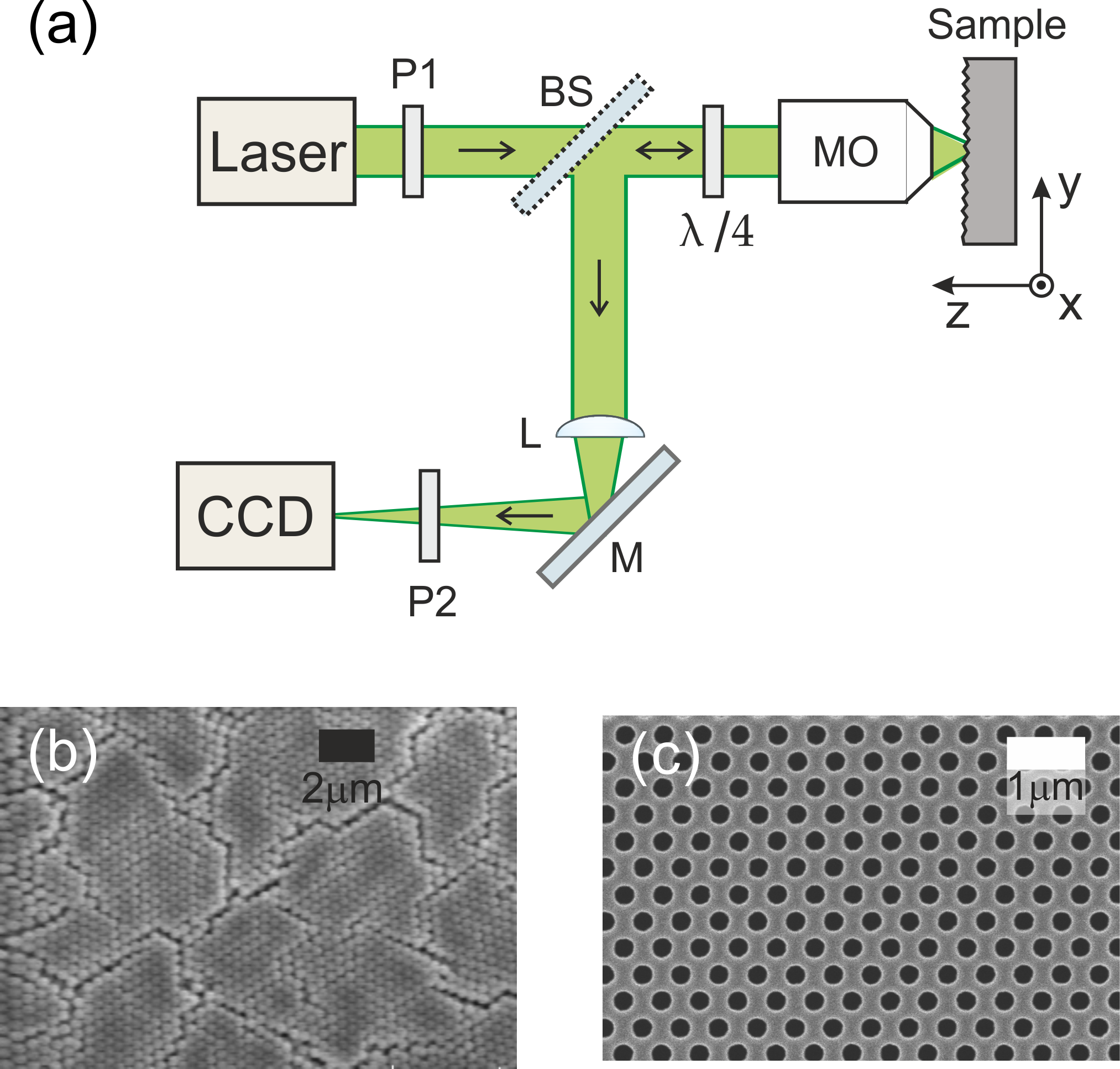}
\caption{\textbf{Experimental setup:} 
A laser beam is focused onto the surface of a sample using a microscope objective (MO). 
The sample is mounted on a three-axis piezoelectric translational stage. 
With a combination of the microscope objective and lens L, the reflected light exiting the front surface of the sample is imaged onto the chip of a charged-coupled device (CCD) camera. 
A polarizer (P2) and a quarter wave-plate ($\lambda/4$) filter the direct reflection from the sample surface.
M -- dielectric mirror, BS -- beam splitter.
The scanning electron microscope image of the (b) fcc (111) top surface of a 3D opal photonic crystal and (c) top view of 2D silicon photonic crystal.}
\label{fig:ch6_setup}
\end{figure}

A schematic of the experimental setup is shown in Fig. \ref{fig:ch6_setup}(a). 
The light source is a continuous wave laser, which emits at a wavelength $\lambda=561\,$nm.
The laser beam passes through a polarizer (P1) and a quarter-wave plate ($\lambda/4$) to give a circularly polarizer light. 
The laser beam is then focused by a microscope objective (MO) (Nikon: Infinity corrected, 100$\times$, NA = 0.9) with a focus diameter of about $300~$nm.
A combination of the microscope objective and lens L (focal length $f = 100~$mm) images the surface of the sample onto the chip of a CCD camera (Dolphin F-145B).
The reflected light was detected in a cross-circular configuration to suppress surface reflections, which was confirmed by a null light reflected from a silicon wafer (a reference mirror). 
In this configuration, a large fraction of the detected light from the sample has been multiply scattered~\cite{OuterThesis,Morgan2003OL}.

In the experiments, we studied light propagation in synthetic opal photonic crystals, which are made of  silica colloidal spheres (radius $R=349\,$nm) grown on a silicon wafer~\cite{Hartsuiker2008Lang,HartsuikerThesis} (see SEM image in Fig.~\ref{fig:ch6_setup}(b)).  
The crystal domains observed as cracks in the SEM image appear during the growth. 
The grain boundaries are a form of disorder that results in multiple scattering.
Other forms of disorder include particle polydispersity and position variation~\cite{Koenderink2005PRB}.
The incident light illuminates the opal with a cone of angles that has its center perpendicular to the fcc (111) plane.
We also performed measurements on a two-dimensional (2D) silicon crystal with centered rectangular lattice as shown in the SEM in Fig.~\ref{fig:ch6_setup} (c). 
The 2D crystal was fabricated by etching pores (radius of $R = 153 \, $nm) using reactive ion etching in a silicon wafer~\cite{WolderingThesis,Woldering2008N}. 
The large single crystal is defined by deep UV step scan lithography~\cite{Woldering2008N}. 
The measurements on 2D crystals are done in the perpendicular out of plane direction with zero in-plane wavevector where the crystal has a stop gap at the $\Gamma$ point. 
In both crystals, the photon energy of the illumination light has a higher energy than the stop gap and couples to high frequency propagating bands.
The transport mean free path of opals have been measured from enhanced backscattering and determined to be of the order of 10 $\,\mu$m~\cite{Koenderink2005PRB,Muskens2011PRB}.
As a reference sample, we used an ensemble of disordered zinc oxide (ZnO) nanoparticles with an average particle size of 200 nm. 
The ZnO disordered sample has a transport mean free path of $\ell = 0.6~\mu$m and a thickness of $L = 10~\mu$m ($L/\ell \approx 17\ell$), therefore the sample is strongly scattering.


\section{Histogram of reflected intensity}

\begin{figure}[ht]
\center
\includegraphics[width=0.45\textwidth]{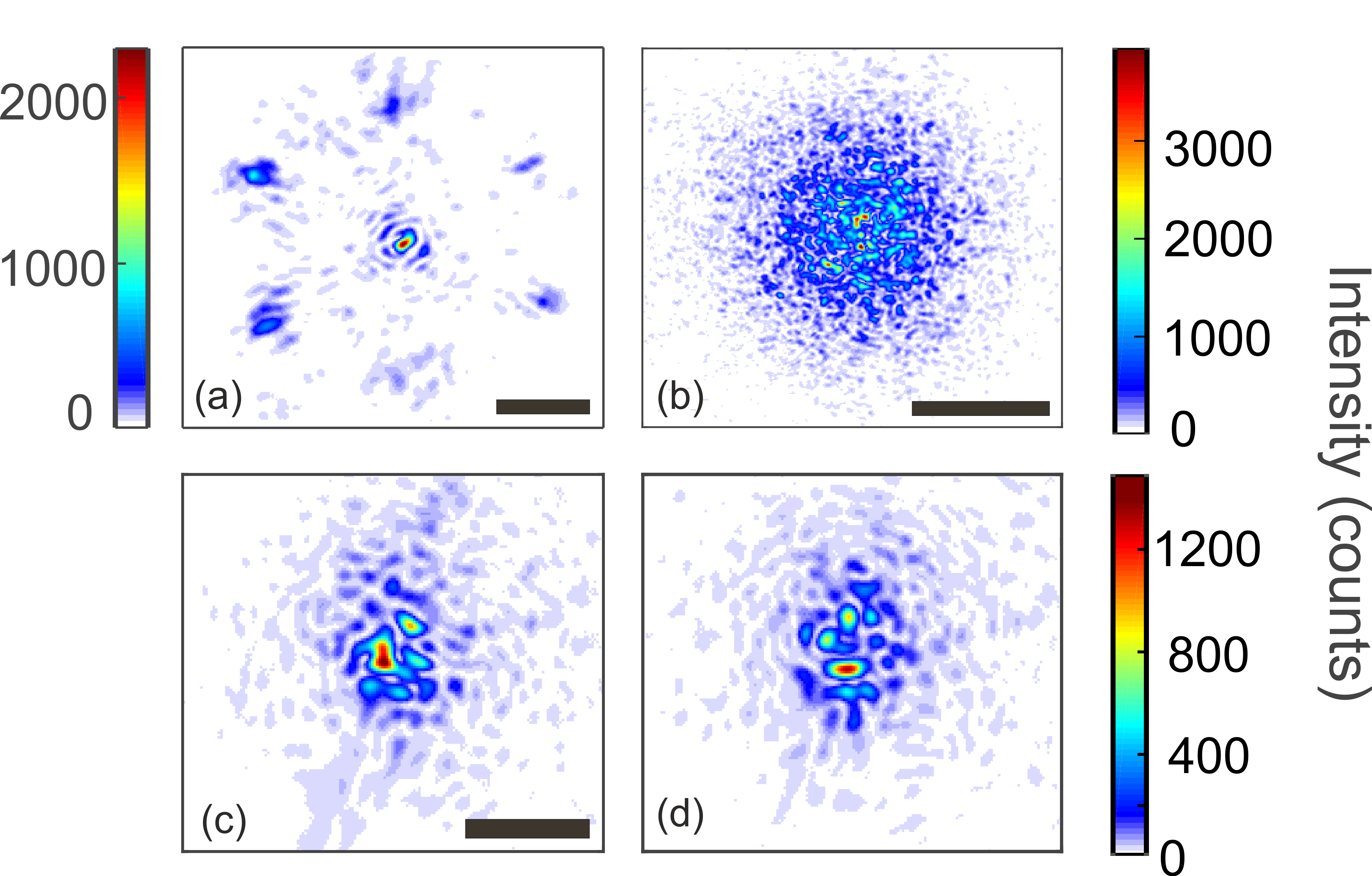}
\caption{CCD camera images of the reflected light from 
(a) opal with a large crystal grain, (b) a disordered ZnO sample, (c) and (d) opal with a small crystal grain, separated by a distance of 100$\,$nm. }
\label{fig:ch6_CCDImages}
\end{figure}

In Fig.~\ref{fig:ch6_CCDImages}(a), we show the reflected intensity from a part of the opal with a large crystal domain. 
The reflected intensity from this part of the crystal has distinct hexagonal peaks.
These hexagonal peaks are attributed to the (111) periodic arrangement of the silica spheres in this part of the crystal and to a large contribution of Bloch waves. 
In contrast to the reflected intensity from a completely disordered ZnO sample shown in Fig.~\ref{fig:ch6_CCDImages}(b), there is no distinct structure due to the effect of random wave interference which results in a speckle pattern.
Figs.~\ref{fig:ch6_CCDImages}(c) and (d) show the images recorded at two different positions on the opal with small crystal grains. 
The reflected intensities from the two positions, separated by a distance of $\Delta x = 100 \,$nm, both have some bright spots, which do not show any hexagonal peaks (see Fig.~\ref{fig:ch6_CCDImages}(a)). 
In this part of the crystal, there is more scattering from the grain boundaries and therefore, the reflected intensity from this part of the crystal is speckle-like.



\begin{figure}[ht]
\center
\includegraphics[width= 0.48\textwidth]{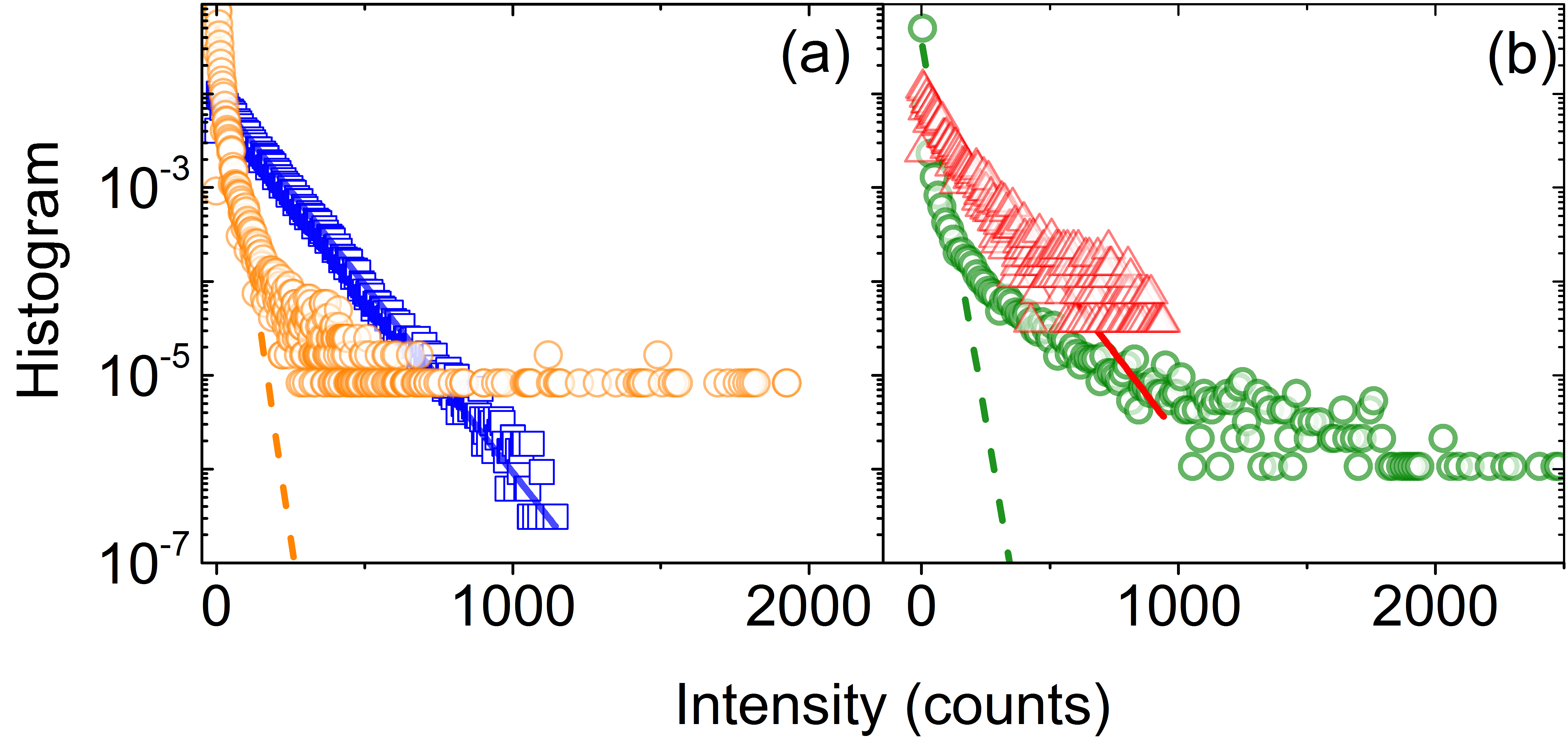}
\caption{ Histogram of speckles intensity reflected from the disordered ZnO sample (blue squares) (a), the opal photonic crystal with large grains (orange spheres) grain (a), the opal with small grains (red triangles) (b), and the 2D silicon photonic crystal (green circles) (b). 
The orange dashed line, blue dash-dotted line, the red solid line, and the green dashed line are the Rayleigh distributions fitted for opal with large crystal grains, ZnO sample, opal with some grains, and 2D silicon photonic crystal, respectively.}
\label{fig:ch6_histogram}
\end{figure}
We analyze the reflected intensity images using histograms, which are plotted in Fig.~\ref{fig:ch6_histogram} for the different samples and are compared with Rayleigh distribution.
The histograms were obtained from the occurrence of intensities within the full-width at half maximum of the beam. 
The occurrence was then normalized to the total occurrence in a single image.

For the ZnO disordered sample, there is a good agreement between the intensity histogram and the Rayleigh distribution, as expected for a disordered sample.
Interestingly, for the large-domain opal crystal, there is a huge deviation of the Rayleigh distribution from the reflected intensity. 
Similarly, we show in Fig.~\ref{fig:ch6_histogram}(b) the probability distribution of intensity reflected from the 2D photonic crystal, which also has a large crystal domain.
For the 2D crystal, there is a strongly non-Rayleigh distribution of intensity as well.
These strong non-Rayleigh distributions imply that the reflected fields are strongly correlated, which is the result of a large contribution of Bloch waves. 
In these large-domain crystals, scattered waves due to disorder have minimal contribution, since the disorder (especially the grain boundaries) hardly affect light propagation.

For the opal photonic crystal with small crystal grains, there is a slight deviation of a negative exponential function from the histogram. 
The deviation from the Rayleigh distribution implies that the reflected intensities are correlated, which is due to the contribution of the Bloch waves. 
At low intensities, there is however an agreement of the Rayleigh distribution with the measured intensity.
We attribute the low intensities to be from the fields that propagate deep inside the sample and have been scattered by disorder.
Therefore, there is a noticeable contribution of the uncorrelated scattered waves due to disorder along side with correlated Bloch waves in the opal with small crystal grains.

\begin{figure}[ht]
\center
\includegraphics[width= 0.6\textwidth]{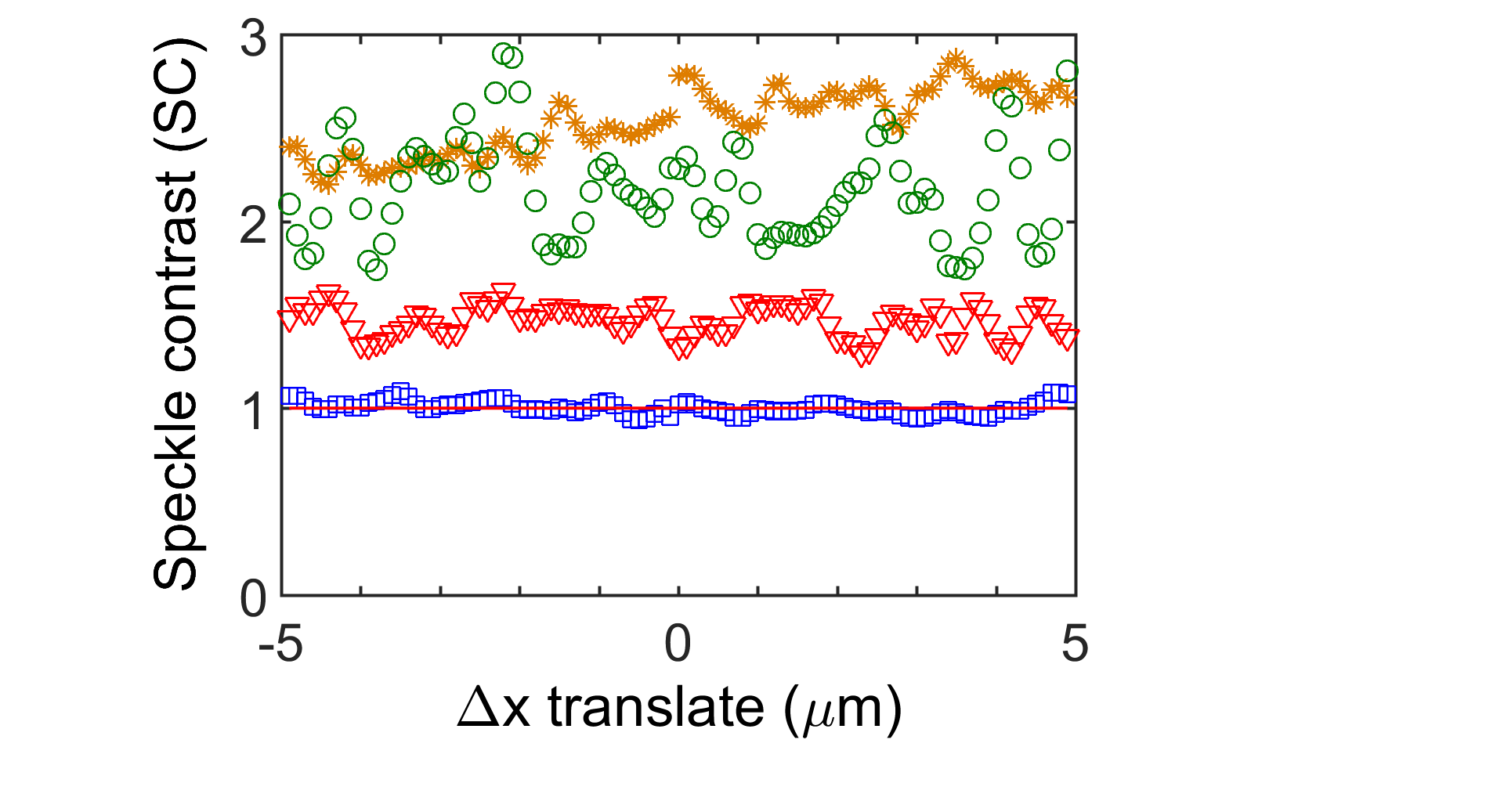}
\caption{Speckle contrast (SC) versus translation along x-axis for a disordered ZnO sample (blue open squares), opal with small grains (red triangles), opal with large grains (orange stars), and large grain 2D silicon crystal (green circles). The red straight line is for speckle contrast SC = 1, which is expected for Rayleigh distribution of intensity.}
\label{fig:ch6_contrast}
\end{figure}

We scanned the sample with a step size of $100~\rm{nm}$ and measured the intensity reflected at each position.
We computed the  speckle contrast (SC) and in Fig.~\ref{fig:ch6_contrast}, we plot the SC versus translation along x axis.
For the disordered sample, the SC fluctuates around 1, which is expected for a completely disordered sample. 
The SC for the photonic crystals however deviates interestingly from 1: the opal with small grain boundaries has a mean SC = 1.3, while the SC for the large-grain crystal shows a gradient. 
This spatial gradient in SC is attributed to the fact that there is a spatial gradient in the disorder strength along the substrate~\cite{Hartsuiker2008Lang}.
The mean SC for the 2D Si crystal is about 2.2 and is comparable to the large-gain opal. 
These comparable SC for these two crystals imply that the amount of disorder in the crystals are comparable.
In general, we observe that the $\rm{SC} > 1$ for all crystals and SC increases with increasing order.
The reason for the increasing SC is due to less amount of scattered waves from disorder and an increase in the contribution of Bloch waves.
We note interestingly that mesoscopic effects get stronger with stronger disorder, however the effect observed here gets stronger with weaker disorder.

\section{Conclusion}
We have employed intensity statistics to study wave transport in synthetic opal photonic crystal and 2D silicon photonic crystal. 
We find that there is a deviation from the Rayleigh distribution, which  we quantified using the speckle contrast. The result depends on the  degree of disorder.
The observed deviations from Rayleigh statistics show that the underlying order plays a significant role in light transport through photonic crystal with disorder. 

\section{Acknowledgments}
We acknowledge Alex Hartsuiker and Leon Wolderink for fabricating the samples and Cock Harteveld for technical assistance. 
This project is part of the research program of the ‘Stichting voor Fundamenteel Onderzoek der Materie’ (FOM) FOM-program ‘Stirring of light!’, which is part of the ‘Nederlandse Organisatie voor Wetenschappelijk Onderzoek’ (NWO). We acknowledge NWO-Vici, DARPA, ERC 279248, and STW.

\end{document}